\documentclass[prb,twocolumn,showpacs,amsmath,amssymb]{revtex4}

\usepackage{graphicx}

\begin{document}

% some shorthand
\newcommand{\refeq}[1]{Eq.\ (\ref{#1})}
\newcommand{\refsec}[1]{Sec.\ \ref{#1}}
\newcommand{\reffig}[1]{Fig.\ \ref{#1}}
\newcommand{\refonline}[1]{Ref.\ \onlinecite{#1}}
\newcommand{\refsonline}[1]{Refs.\ \onlinecite{#1}}

\newcommand{\ket}[1]{\hbox{$\mid \! {#1} \rangle$}}
\newcommand{\bra}[1]{\langle #1 |}
\newcommand{\braket}[2]{\langle #1 | #2 \rangle}
\newcommand{\bigbraket}[3]{\langle \, #1 \, | \, #2 \, | \, #3 \, \rangle}
\newcommand{\com}[2]{{\big[} #1 \, , \, #2 {\big]}}
\newcommand{\anticom}[2]{{\big\{} #1 \, , \, #2 {\big\}}}
\newcommand{\ie}{\textit{ie.\ }}
\newcommand{\Tr}{\mathop{\mathrm{Tr}}}

\title{Infinite size density matrix renormalization group, revisited}
\author{I.P. McCulloch}
\affiliation{Institut f\"ur Theoretische Physik C, RWTH Aachen University, D-52056 Aachen, Germany}
\affiliation{School of Physical Sciences, The University of Queensland, Brisbane, Queensland, 4072, Australia}
%\pacs

\begin{abstract}
I revisit the infinite-size variant of the Density Matrix Renormalization Group (iDMRG)
algorithm for obtaining a fixed-point translationally invariant matrix product 
wavefunction in the context of one-dimensional
quantum systems. A crucial ingredient of this algorithm is an efficient transformation
for obtaining the matrix elements of the wavefunction as the lattice size is increased, and I introduce
here a versatile transformation that is demonstrated to be much more effective than previous versions.
The resulting algorithm has a surprisingly close relationship to Vidal's Time Evolving Block Decimation 
for infinite systems, but allows much faster convergence.
Access to a translationally invariant matrix product state allows the calculation of
correlation functions based on the transfer matrix, which directly gives the spectrum
of all correlation lengths. I also show some advantages of the Matrix Product Operator (MPO) technique
for constructing expectation values of higher moments, such as the exact variance 
$\langle (H-E)^2 \rangle$.
\end{abstract}

\maketitle

\section{Introduction}

For many years, the density-matrix renormalization-group algorithm 
\cite{White,UliReview} (DMRG) has been used in an increasing diversity of applications
in fields such as condensed matter physics,
condensed matter physics, quantum chemistry, nuclear physics and quantum information science. 
The DMRG algorithm is used to obtain a variational approximation for groundstates and
low-lying excited states, with later variants being developed for finite-temperature and
dynamical properties. The basic algorithm comes in two varieties.
The most commonly used algorithm is the so-called \textit{finite-size} algorithm,
whereby a matrix-product state (MPS) wavefunction on a finite-size lattice
(defined through a distinct set of matrices $A_n^{s_n}$, for each lattice site
$n = 1,2,\ldots,L$, where $\ket{s_n}$ is a $d$-dimensional basis for the local Hilbert space at site $n$) 
is iteratively improved until convergence. The \textit{infinite-size} algorithm,
on the other hand, grows the lattice by one or more sites each iteration and produces,
at the fixed point, a wavefunction that is translationally invariant (possibly with a non-trivial
unit cell). This approach is useful for probing quantities in the thermodynamic limit, without
the influence of boundary effects,
% \footnote{Note however in some cases the effect of a restricted
%basis size in an MPS can have a similar effect to a boundary, for example in spontaneously breaking
%translation symmetry.} 
and such a translationally invariant state can readily be used as an initial state to 
study real-time evolution of a homogeneous quench as well as local (translation symmetry breaking) 
perturbations. 
Recently, interest in MPS algorithms that obtain a thermodynamic fixed point has been rekindled with
the invention of a variant of Vidal's Time-Evolving Block Decimation (TEBD) algorithm \cite{TEBD}
that applies directly to an infinite size wavefunction to give iTEBD
\cite{iTEBD}, the resulting wavefunction being translationally invariant by construction
(although
with the restriction that the unit cell must be at least two sites in size).
The iTEBD utilizes imaginary time evolution via a Trotter-Suzuki decomposition \cite{TrotterSuzuki},
amounting to using the power method \cite{GolubVanLoan} to obtain the groundstate eigenvector of
an approximation of the exponential of the Hamiltonian. However the power method is not an efficient
eigensolver, and the use of the Trotter-Suzuki decomposition requires careful scaling of the 
time-step to zero in order to obtain a properly converged state. Hence, all else being
equal, a DMRG approach where an efficient local eigensolver is used to find a variationally optimal
state ought to be much more robust and efficient.

In the past, the main use for the infinite size DMRG algorithm however has not been to achieve
a fixed point, but rather to grow an initial wavefunction of a particular size as an initialization
step for the finite-size algorithm \cite{White}. As an algorithm for obtaining a translationally
invariant fixed point, infinite-size DMRG (hereafter referred to as iDMRG) has been virtually abandoned.
There are several reasons for this: Firstly, finite-size DMRG (fDMRG) readily allows
the application finite-size scaling \cite{FiniteSizeScaling} with respect
to the lattice size, whereas for iDMRG the
only available parameter is the number of states kept $m$. In fact, this is not a problem
because the scaling relations in terms of the number of states have already been investigated
by several authors
\cite{NishinoScaling,Boman,Luca}, 
%leading to scaling relations for the correlation length
%and other quantities in terms of $m$,
although having received insufficient attention this
approach is not yet widely utilized. In fact, having only a single scaling parameter is a marked
advantage of iDMRG over finite size approaches where either the number of states kept must be sufficiently
large that the truncation effects are negligible, which may be impractical, 
or $m$ must be considered as an additional
parameter and scaling performed with respect to both parameters, which in the past has not always
been done correctly (see \refonline{NishinoScaling} for the correct approach).

A second disadvantage of iDMRG over fDMRG is a technical limitation of the algorithm,
namely the lack of a good initial guess vector for the 
local eigensolver at each DMRG iteration. In contrast, 
the development of the \textit{wavefunction acceleration}
procedure in fDMRG \cite{WhiteAccel}, 
to transform the basis for the superblock wavefunction from one iteration to be the
initial guess vector for the next iteration, was a breakthrough in the development of 
finite-size DMRG
that improves the efficiency of the algorithm by orders of magnitude. This transformation
is essentially implicit in the matrix product formulation of DMRG\cite{Rommer,VerstraetePBC,DMRGMPS}.
From this point of view, DMRG is an algorithm for locally updating the matrix elements of an MPS
one (or a few) lattice sites at a time, and the wavefunction transformation results from the
orthonormalization the basis which is necessary to ensure numerical stability. 
However, this transformation only applies to finite-size MPS, and the apparent lack of a similar 
transformation for iDMRG has been the focus of several studies in the last decade or more.

The problem that needs to be solved to obtain a wavefunction transformation for iDMRG is to
% fix the sign of the basis elements, and to
find a procedure for predicting the sign of the new basis elements
when the lattice size is increased. This sign structure has two components, being the intrinsic
nodal-structure of the wavefunction, as well as a phase ambiguity of the eigenvectors
of the density matrix (or singular value decomposition). There have been several works that attempt
to obtain a transformation algorithm, either using Marshall's sign rule \cite{Marshall} to predict
the sign structure of the wavefunction \cite{MarshallDMRG}, or directly comparing the sign structure
from different steps \cite{DMRG-Qin,DMRG-Sun}. These approaches have had mixed success,
in that they rely on a $1-1$ matching of density matrix eigenvalues (equivalently, singular
values arising from the Schmidt coefficients) from successive iterations and
attempting to match signs of the eigenvectors (say, by choosing that the first component of the vector
will always be positive). Where applicable, the Marshall's sign rule approach is very effective \cite{MarshallDMRG},
but the need to know the nodal structure of
the wavefunction \textit{a priori} is a severe limitation, eg.~this
approach breaks down completely for frustrated systems.
The approach of Qin and Lou does lead to a wavefunction that is somewhat better
than a random state, and in this sense this approach is useful in reducing the number
of iterations required by the local eigensolver. However, repairing eigenvector signs
in this manner is not robust and this procedure has not been demonstrated to be effective in obtaining a
fixed point wavefunction; that is, in the studies utilizing this approach \cite{DMRG-Qin,DMRG-Sun} the
algorithm does not converge to a point where the transformation
reproduces the fixed point wavefunction with full fidelity.

It may be a surprise, therefore, that a solution to most of these problems exists in the literature 
and has done for quite some time, pre-dating even the wavefunction transformation in fDMRG, under
the name of Product Wave-Function Renormalization Group (PWFRG) \cite{PWFRG,PWFRG2,PWFRG3,PWFRG-MPS}.
% \textbf{More about PWFRG,
%do we need to describe the algorithm?  If so it should have a separate section.}.
In this approach, a recurrence relation is set up to obtain the translation operator that shifts
the wavefunction by one site (or unit cell, which may be an arbitrary number of lattice sites). 
Using this operator the initial wavefunction for the next iteration
can be obtained. However,
while this procedure is effective in obtaining a translationally invariant fixed point, it still
relies on matching singular values from successive iterations and the rate of convergence \textit{away}
from the fixed point is quite poor. This is because
the wavefunction transformation does not have full fidelity until the 
translation operator itself has converged,
however the translation operator obtained from the recurrence relation converges is quite slowly.
A main result of this paper is the introduction of a new
transformation that reaches the same fixed point as PWFRG, but converges much faster when away from
the fixed point, \ie when the lattice size is small. In fact, the convergence is
effective enough that this approach is useful even for exact diagonalization of small systems.
Also in this paper, I investigate some approaches to calculating correlation functions using
the transfer operator. This approach has been known for quite some time \cite{Rommer,RommerOstlund,Boman}, 
but has
not been widely utilized. Thirdly, the more recent development of the Matrix Product Operator (MPO)
formulation \cite{DMRGMPS}, for obtaining the exact matrix product representation of the 
Hamiltonian and observables is demonstrated to have significant advantages and I present
algorithms for obtaining expectation values in the infinite-size limit 
directly from the MPO representation.

The layout of this paper is as follows:
In section \ref{sec:MPS}, I describe the basic matrix product approach and notation,
for both wavefunctions and operators.
In section \ref{sec:iDMRG} I describe the iDMRG algorithm itself and some discussion of
the convergence properties. Section \ref{sec:Expectation} treats the calculation of observables
for many cases including local quantities, correlation functions including the spectrum of
correlation lengths, groundstate fidelities,
and translationally-invariant operators such as the energy and the variance.
Finally, section \ref{sec:Conclusions} contains a summary and some concluding remarks.

\section{Matrix product states}
\label{sec:MPS}

We denote an MPS on an $L$-site lattice by the form,
\begin{equation}
\Tr \sum_{\{s_i\}} A^{s_1} A^{s_2} \cdots A^{s_L}
\quad \ket{s_1} \otimes \ket{s_2} \otimes \cdots \otimes \ket{s_L} \; .
\label{eq:MPWavefunction}
\end{equation}
The local index $s_i$ represents an element of the $d$-dimensional local Hilbert space at site $i$,
and the $A^{s_i}$ are $m \times m$ matrices, $m$ being the dimension of the 
\textit{matrix basis} (this quantity is also often called $D$, and sometimes $\chi$).
As a short-hand notation, it is convenient to omit the basis states $\ket{s_i}$ and the summation from 
\refeq{eq:MPWavefunction} and speak of an MPS as being defined purely as a sequence of
matrices. This notation is quite consistent if one regards the local index $s_i$ as being a ket vector.
Thus, in shorthand notation, \refeq{eq:MPWavefunction} is equivalent to simply
$A^{s_1} A^{s_2} \cdots A^{s_L}$.

In practice, a MPS state with no particular constraints on the
form of the $A$-matrices is numerically difficult to handle and it is usual to impose
some orthonormalization constraints (see eg.~\refonline{DMRGMPS} for details),
\begin{equation}
\sum_s A^{s\dagger} A^s = 1 \; ,
\label{eq:LeftOrtho}
\end{equation}
for the left-orthonormalization constraint, and conversely
\begin{equation}
\sum_s B^s B^{s\dagger} = 1 \; ,
\label{eq:RightOrtho}
\end{equation}
for the right-orthonormality constraint. In this paper, I use always $A^s$ to denote
a set of matrices satisfying the left-orthonormalization constraint \refeq{eq:LeftOrtho},
and $B^s$ to denote a set of matrices satisfying the right constraint \refeq{eq:RightOrtho}.
The infinite-size DMRG algorithm is based around adding sites to the center of the chain.
To this end, it is useful to orthonormalize the sites of the MPS from the edges, working
in towards the center, to give a wavefunction of the form,
\begin{equation}
A^{s_1} \ldots A^{s_{n-1}} A^{s_n} \Lambda B^{s_{n+1}} B^{s_{n+2}} \ldots B^{s_L}\; .
\label{eq:MPSWavefunctionCenter}
\end{equation}
Here $\Lambda$ is a matrix that is left over after all of the sites in the MPS have been
orthonormalized, which I refer to as the \textit{center matrix}, and this matrix
is also equivalent to what is called the \textit{superblock wavefunction} in DMRG terminology. \cite{UliReview}
In \refonline{DMRGMPS}, I used $C$ for this matrix, here I use $\Lambda$ to stand for the
matrix of singular values $(\lambda_i)$ as start towards harmonizing the various notations
associated with DMRG and MPS approaches.

These orthonormalization conditions correspond to a basis in which the reduced density matrix for the left (right) half of the
system is diagonal, and are imposed for example by performing a singular value decomposition treating
$A^s$ as a single $dm \times m$ matrix (or $B^s$ as a $m \times dm$ matrix),
\begin{equation}
\begin{array}{rcl}
A &=& \left[ \begin{array}{l} A^1 \\ A^2 \\ \vdots \\ A^d \end{array} \right] \vspace{3mm} \\
B &=& \left[ B^1 B^2 \ldots B^d \right]
\end{array}
\label{eq:ABDecomp}
\end{equation}
and then performing the singular value decomposition $A = U \Lambda V^\dagger$, 
where $\Lambda$ is a $m \times m$ diagonal matrix such that $\Lambda^2$ is the reduced density
matrix, and $U,V$ are row-unitary matrices. Upon rewriting $U$
back as a set of $m \times m$ matrices
analogously to \refeq{eq:ABDecomp}, one obtains $U^s$ as satisfying the orthogonality constraint \refeq{eq:LeftOrtho},
with a remainder term $\Lambda V^\dagger$ appearing on the right (which we can, for example,
incorporate into the neighboring $A$-matrix). The procedure for obtaining the right-orthonormalization constraint
is similar, now ending up with a remainder matrix appearing on the left.
In this fashion, the boundary of the left/right orthonormalized states can be moved to an arbitrary position in the
lattice, a procedure I refer to as a \textit{rotation} of the center matrix to the left or right.

The infinite-size DMRG produces is a wavefunction that grows by adding sites in the center,
which, for the example case of a 2-site unit cell, gives a wavefunction at the $n^\mathrm{tm}$ iteration
that is $2n$ sites long,
\begin{equation}
A_1^{s'_1} \ldots A_{n-1}^{s'_{n-1}} A_n^{s'_n} \Lambda_n B_n^{s_{n}} B_{n-1}^{s_{n-1}} \ldots \; B^{s_1}_1 \;.
\label{eq:MPSGrowing}
\end{equation}
Sites are added until the the $A^s$ and $B^s$ matrices in the center of the lattice converge to a fixed point. 
At this fixed point, the MPS will be translationally invariant, possibly with a non-trivial unit cell,
such that the $A^s,B^s$ matrices are independent of position. In TEBD the associated MPS
is always of this translationally invariant form, but in iDMRG we will see that the translational invariance arises from
the wavefunction transformation converging to the identity operator at the fixed point.

The MPS form \refeq{eq:MPWavefunction}  is related to the canonical form used by Vidal \cite{TEBD},
\begin{equation}
\ldots \Gamma^{s_{n-1}} \Lambda \Gamma^{s_n} \Lambda \Gamma^{s_{n+1}} 
\Lambda \Gamma^{s_{n+2}} \ldots \; ,
\end{equation}
by $A^s = \Lambda \Gamma^s$ and $B^s = \Gamma^s \Lambda^s$. For the canonical form, the
orthogonality constraints take the form
\begin{equation}
\begin{array}{rcl}
\sum_s \Gamma^{s\dagger} \rho_R \Gamma^s &=& 1 \\
\sum_s \Gamma^{s} \rho_L \Gamma^{s\dagger} &=& 1 \; ,
\end{array}
\end{equation}
where $\rho_L = \Lambda \Lambda^\dagger$ and $\rho_R = \Lambda^\dagger \Lambda$ are the reduced density
matrices for the left and right semi-infinite strips respectively.
This shows that the $\Gamma^s$ and $\Lambda$ matrices for a canonical MPS are not independent.
In terms of the $A^s,B^s$ matrices, the corresponding fixed point relations are, \cite{MPSRepresentations}
\begin{equation}
\begin{array}{rcl}
\sum_s A^{s} \rho_L A^{s\dagger} &=& \rho_L \\
\sum_s B^{s\dagger} \rho_R B^s &=& \rho_R \; .
\end{array}
\label{eq:FixedPointDM}
\end{equation}

States that have a non-zero quantum number can be readily obtained in this scheme, as long
as a repeated unit cell can be properly defined.
For example, for finite doping or finite magnetization with a density $p/q$ for coprime integers $p,q$, we use a 
unit cell of $q$ sites and insert a shift in the quantum numbers of $-p$ per unit cell.
The effect of this shift in the wavefunction (equivalent to an auxiliary particle
of quantum number $-p$) is to force the physical system to have an average quantum number
of $+p$ throughout the unit cell, while the entire MPS (physical system plus the quantum number shift) 
has an average quantum number of zero and thereby allowing a repeated unit cell.
This shift can be implemented in a variety of equivalent ways, for example by
introducing explicitly an additional matrix $Q$ into the MPS unit cell that
has matrix elements $\bigbraket{j'}{Q}{j} = \delta_{j',j-p}$, or by shifting the quantum numbers
in the local basis of one site in the unit cell by $-p$.
Note that this procedure does not enforce that there will be \textit{exactly} $p$ particles
per unit cell, which would be an undesirable restriction, but arbitrarily large fluctuations
are allowed within the constraint that the average quantum number on the \textit{infinite} system is $p/q$.

\subsection{Matrix product operators}

\refonline{DMRGMPS} introduces a representation for the Hamiltonian and other 
associated operators as an exact matrix product operator (MPO). These operators come in several
varieties, the main form that we consider here is operators that are
characterized by an MPO representation
that can be put in a lower (upper) \textit{triangular} form, and therefore represent operators 
that are polynomial combinations of local operators, hence the expectation value of such an operator
is a polynomial function of the lattice size, and in typical cases will be linear, for example
giving the energy per site.
A second form of MPO arises from products of local operators, such as occurs in the exponential
of a triangular operator. This form has an expectation value that is an exponential function $\lambda^L$
of the lattice size, hence I call this class \textit{exponential MPO's}.
Crosswhite and Bacon \cite{Crosswhite} have also considered MPO's from the point of view of 
a correspondence with weighted finite-state automata.

Firstly, we briefly examine some properties of triangular MPO's. 
The notation we use here is that an MPO is a matrix of local operators of dimension $M \times M$.
This form, after becoming familiar with the notation, is a very convenient way of representing an MPO,
see \refonline{DMRGMPS} for full details.
For simplicity we take here the unit cell to be a single site, so that the infinite MPO
is represented by the same matrix repeated on every site.
For cases where the unit cell of the Hamiltonian is non-trivial, for example an $N$-leg ladder,
the associated MPO has the same periodicity which leads to the technicality of
treating MPO matrices that may be non-square (trapezoidal, in the case of triangular MPO's), but
this presents no serious difficulty and all of the procedures described here carry over
to this case straightforwardly.
Starting from the most general lower-triangular $2 \times 2$ MPO,
\begin{equation}
W = \begin{pmatrix}
C & 0 \\
B & A
\end{pmatrix} \; ,
\end{equation}
this represents the infinite sum of all terms of the form
\begin{equation}
A \otimes \cdots \otimes A \otimes B \otimes C \otimes \cdots \otimes C \; .
\end{equation}
Note that \refonline{DMRGMPS} used $M$ to denote the MPO matrix, whereas here I use $W$
to avoid confusion with $M$, the dimension of the $W$ matrix.
By appropriate choices of local operators $A,B,C$, this form can be used for many variants of
one-particle operators, including
bosonic and fermionic particle operators at arbitrary lattice momenta, or the sum of local interactions.
Moving to $3 \times 3$ matrices, we can generalize this form to a bond operator, or by including a term on
the diagonal, 
\begin{equation}
W = \begin{pmatrix}
E & 0 & 0 \\
D & C & 0 \\
0 & B & A
\end{pmatrix} \; ,
\end{equation}
which represents the sum of all string-like terms of the form
\begin{equation}
A \otimes \cdots \otimes A \otimes B \otimes C \otimes \cdots \otimes C 
\otimes D \otimes E \otimes \cdots \otimes E\; .
\end{equation}
Thus long-range string operators can be represented just as easily as short-range terms.
Taking the string operator $C$ to be proportional to the identity operator shows that 
non-local but exponentially decaying interactions can be used in DMRG with negligible
loss of efficiency over purely local interactions, a fact that is probably well known
to DMRG practitioners although I am not aware of any studies that make use of this.

For a concrete example of a triangular MPO, combining a bond term and a local term 
a simple but non-trivial Hamiltonian operator is the Ising model in a transverse field (ITF),
\begin{equation}
H = \sum_{<i,j>} \sigma^z_i \sigma^z_j + \lambda \sum_i \sigma^x_i \; ,
\label{eq:ITF}
\end{equation}
which has the lower-triangular MPO form as $3 \times 3$ matrices,
\begin{equation}
W = \begin{pmatrix} 
1 & 0 & 0 \\ 
\sigma^z & 0 & 0 \\
\lambda \sigma^x & \sigma^z & 1
\end{pmatrix} \; .
\end{equation}

The use of MPO's gives a convenient way to construct a software code
that has no explicit dependence on the particular Hamiltonian. 
The main advantage, however, is that the MPO
form allows arithmetic operations, in particular sums and products of operators 
are represented
simply by taking the matrix direct sum and direct product respectively 
of the MPO representation,
as described in \refonline{DMRGMPS}. 
If both operators have
a lower-triangular MPO representation, then the sum or product is also a 
lower-triangular MPO.
This means that it is very easy to obtain operators that would take
considerable effort to construct using the ad-hoc methods traditionally used in 
DMRG and MPS approaches.
One such example is the variance, $(H-E)^2$, which replaces the discarded weight
(the \textit{truncation error}, in DMRG terminology) 
as the convergence measure of choice
as it gives a measure that is both independent of the details of the 
algorithm and a more reliable
indicator of convergence. 

Not all useful operators have the form of a triangular MPO. The triangular MPO's represent
sums of either short-range or string terms, whereas other kinds of operators of interest
comprise instead products of terms. An example of this is the MPO
representation of the Trotter-Suzuki approximation of an exponential, for example
the first order decomposition of the exponential of a Hamiltonian operator with 
only nearest-neighbor terms into terms acting on even and odd bonds,
$\exp[-\beta H] \simeq \exp[-\beta H_{\mathrm{even}}] \exp[-\beta H_{\mathrm{odd}}]$.
The MPO representation of $\exp[-\beta H_{\mathrm{even}}]$ has a 2-site unit cell with
alternating $1\times M$ and $M \times 1$-dimensional $W$-matrices representing
the product of all of the exponentials for the even bonds, and similarly
$\exp[-\beta H_{\mathrm{odd}}]$ alternates $M \times 1$ and $1\times M$ matrices. The product of the
two operators therefore is an $M \times M$ MPO that will not be lower-triangular.
This operator falls into the class of exponential MPO's, with
an expectation value of the form $\lambda^L$ for $L$ sites,
where $\lambda$ is the eigenvalue corresponding to the fixed point $E^\alpha$ eigenmatrices (see below).

The MPO formulation gives a natural form for obtaining the actual matrix elements of an operator
in the form required to calculate an observable using an MPS wavefunction.
This is done by obtaining matrices $E^\alpha$ and $F^\beta$ that represent the operator
matrix elements on the left and right halves of the system respectively.
Later we will see how to obtain these matrices for infinite states directly, however let us first
consider how to construct these matrices for a finite, but arbitrary size system, which precisely what is
needed for the iDMRG algorithm.
Let us take the wavefunction of size $2n$ sites, given by \refeq{eq:MPSGrowing}.
The initial $E^\alpha,F^\beta$ matrices for the boundaries are taken to be
\begin{equation}
\begin{array}{rcl}
E_0^\alpha &=& \delta_{\alpha,M} I_{\mathrm{left}} \; , \\
F_0^\beta &=& \delta_{\beta,1} I_{\mathrm{right}} \; . \\
\end{array}
\end{equation}
where $I_{\mathrm{left}}$ and $I_{\mathrm{right}}$ are the identity operators acting on the left- and right-hand edges of
the wavefunction. For an open boundary condition state, which is always what we use in iDMRG, these operators
will simply be $1\times 1$ identity matrices.
The $E^\alpha,F^\beta$ for the remainder of the chain are then defined recursively,
\begin{equation}
\begin{array}{rcl}
E^{\alpha'}_n &=& \sum_{s',s,\alpha} W^{s's}_{\alpha'\alpha} 
A^{s'_n\dagger} E^{\alpha}_{n-1} A^{s_n} \\
F^{\alpha'}_n &=& \sum_{s',s,\alpha} W^{s's}_{\alpha'\alpha}
B^{s'_n} F^{\alpha}_{n-1} B^{s_n\dagger}
\end{array}
\end{equation}
Once we have obtained the matrices $E_n^\alpha,F_n^\beta$ for the center of the chain,
we can perform many operations. For example, the action of the operator on the center matrix
in its projected Hilbert space is 
\begin{equation}
W(\Lambda_n) = \sum_{\alpha} E_n^\alpha \Lambda_n F^{\beta\dagger}_n \; ,
\label{eq:ActionCenter}
\end{equation}
and the expectation value is given by the matrix inner product,
\begin{equation}
\bigbraket{\Lambda_n}{W}{\Lambda_n} = \Tr E_n^\alpha \Lambda_n F^{\beta\dagger}_n \Lambda^\dagger \; .
\label{eq:MPSExpectation}
\end{equation}

Note that the action on the center matrix \refeq{eq:ActionCenter} occurs in the \textit{projected} Hilbert space
of the $m\times m$-dimensional basis that is the support of $\Lambda_n$, and  \refeq{eq:ActionCenter} in no way represents
the action of the operator on the entire state (but that action is easy to obtain, see for example \refonline{DMRGMPS}).
It would be a serious error to expect, for example, that the action of the product of two operators $W_1 W_2$
on a wavefunction can be obtained by applying separately the $E,F$ matrices for $W_2$ and then the $E,F$ matrices
for $W_1$. That action would correspond instead to the operator $W_1 P W_2$, where $P$ is the projector onto
the $m\times m$-dimensional Hilbert space. 
Nevertheless, for a properly constructed set of $E,F$ matrices,
the expectation value \refeq{eq:MPSExpectation} is \textit{exact}, because it results from the inner product
of the action of the operator with the original state, for which the fact that this operation 
is being performed within a projected Hilbert space does not matter.

%\subsection{Local updates}

Having seen how to obtain matrix elements in the projected Hilbert space of a $\Lambda$ matrix in the MPS, the final
ingredient before presenting the iDMRG algorithm is to see how the MPS itself is iteratively updated.
For this task, we need the action of an operator on not just a $\Lambda$ matrix, but on
one (or more) of the $A^s$, $B^s$ themselves. This is most easily demonstrated for the action of an MPO on
a single site and incorporating the center matrix. Thus, taking the wavefunction of \refeq{eq:MPSGrowing}
and leaving all matrices fixed except $A_n^{s'_n}$ and $\Lambda_n$, we obtain the action of an MPO on these matrices,
\begin{equation}
W(A_n^{s'_n} \Lambda_n) = \sum_{\alpha,\beta,\sigma} E^\alpha_{n-1}  \;
A_n^{\sigma} \Lambda_n \; F^{\beta \dagger}_n \; W^{\sigma s'_n}_{\alpha\beta} \; ,
\end{equation}
where we have used the explicit matrix elements of the $W$ matrix defining the MPO $W^{\sigma s'_n}_{\alpha\beta}$, with
$\sigma, s'_n$ indices running over the $d$-dimensional local basis and $\alpha,\beta$ indices running over the $M$-dimensional matrix basis
of the MPO. For modifying two or more sites at a time, analogous formulas apply with one $W$ matrix appearing per site.

\section{iDMRG}
\label{sec:iDMRG}

In this section, I describe the iDMRG algorithm and discuss some convergence and stability issues.
The algorithm itself is presented in figure \ref{fig:iDMRG}.
For simplicity I have used a unit cell of two sites, however the unit cell can be
any size, including a single site.
% \textbf{are the modifications required for a single-site
%unit cell obvious to the reader?} 
If the system is parity-symmetric then
step 3 can be avoided with
the $B^{s}$ matrices taken to be the transpose of $A^s$ (if good quantum numbers are used,
the quantum numbers in the matrix basis must be flipped as well, otherwise one ends up with
a conjugate tensor rather than a normal tensor \cite{DMRGMPS}), being careful to use an eigenvalue
decomposition rather than a singular value decomposition 
to ensure symmetry of the left and right basis vectors.
For unit cells larger than two sites, the only modification
is steps 2 and 3 rotate the center matrix through the remainder of the unit cell, optimizing
the matrices comprising the unit cell in the process. Thus, this
transformation gives a recurrence relation whereby a wavefunction on an $n$-site lattice
and a wavefunction on a larger $n+k$-site lattice for some arbitrary $k>0$,
are used to obtain an approximate wavefunction for an $n+2k$-site lattice.
Note that steps 2 and 3 act independently on $\ket{\Psi_n}$ and could be performed in parallel.

\begin{figure}[ht]
\begin{tabular}{cp{0.8\columnwidth}}
\small
1. & (initialization) Obtain the wavefunction for a two-site lattice
$\ket{\Psi_0} = A_{0}^{s'_{0}} \Lambda_{0} B_{0}^{s_{0}}$ and a four-site lattice
$\ket{\Psi_1} = A_{0}^{s'_{0}} A_{1}^{s'_{1}} \Lambda_{1} B_{1}^{s_{1}} B_{0}^{s_{0}}$,
and set $n=1$ for the start of iterations. \\

2. & Rotate the center matrix of $\ket{\Psi_n}$ one step to the left, to obtain
$\ldots \Lambda^L_n B_{n+1}^{s'_{n+1}} B_n^{s_{n}} \ldots$. \\

3. & Rotate the center matrix of $\ket{\Psi_n}$ one step to the right, to obtain
$\ldots A_n^{s'_{n}} A_{n+1}^{s^{}_{n+1}} \Lambda^R_n \ldots$.\\

4. & (new wavefunction) The trial wavefunction for a lattice with size increased by two sites is given
by $\ket{\Psi^{\mathrm{trial}}_{n+1}} = \ldots A_{n+1}^{s'_{n+1}} 
\Lambda^R_n \Lambda_{n-1}^{-1} \Lambda^L_n B_{n+1}^{s_{n+1}} \ldots $.\\

5. & (eigensolver) Use $\ket{\Psi^{\mathrm{trial}}_{n+1}}$ as the initial guess
vector for an eigensolver, updating the matrices $A_{n+1}^{s'_{n+1}} \Lambda_{n+1} B_{n+1}^{s_{n+1}}$
to obtain the final $\ket{\Psi_{n+1}}$.\\

6. & (truncation) Truncate the dimension of $\Lambda_{n+1}$ as desired, using eg.~ a singular
value decomposition or eigenvalue decomposition, and normalize.\\

7. & (stopping criteria) If $\Lambda_{n+1}$ is sufficiently close to $\Lambda^{L,R}_{n}$, then
the fixed point has been reached. Otherwise increase $n \leftarrow n+1$ and return to step 2.\\
\end{tabular}
\caption{The iDMRG algorithm, including the wavefunction transformation.}
\label{fig:iDMRG}
\end{figure}

The key component of this algorithm is step 4, where the wavefunction center matrix 
$\Lambda_{n+1}^{\mathrm{trial}}$ is obtained from those of the previous two iterations,
\begin{equation}
\Lambda_{n+1}^{\mathrm{trial}} = \Lambda^R_n \Lambda_{n-1}^{-1} \Lambda^L_n \; .
\label{eq:iDMRGTransformation}
\end{equation}
Note that $\Lambda^L_n$ and $\Lambda^R_n$ are center matrices of the same wavefunction $\ket{\Psi_n}$,
but located at different partitions, at the left and right boundaries of the unit cell respectively.
\refeq{eq:iDMRGTransformation}
is a straightforward computation that, unlike the PWFRG transformation \cite{PWFRG} and sign-rule
based approaches, can be applied directly without any kind of matching of singular values from different
basis sets. This is
because the basis for the right hand side of $\Lambda^R_n$ is the same as the right hand side of $\Lambda_{n-1}$,
and similarly the left hand side of $\Lambda^L_n$ matches the left hand side of $\Lambda_{n-1}$.
Thus there is not even any requirement here to diagonalize the $\Lambda$ matrices, they need not even be square.
For simplicity of calculating the inverse $\Lambda_{n-1}^{-1}$ however, one can choose a basis in which
$\Lambda_{n-1}$ is diagonal, which is normally obtained as part of the truncation step anyway.

\begin{figure}[ht]
\centering
\includegraphics[width=\columnwidth,clip=true]{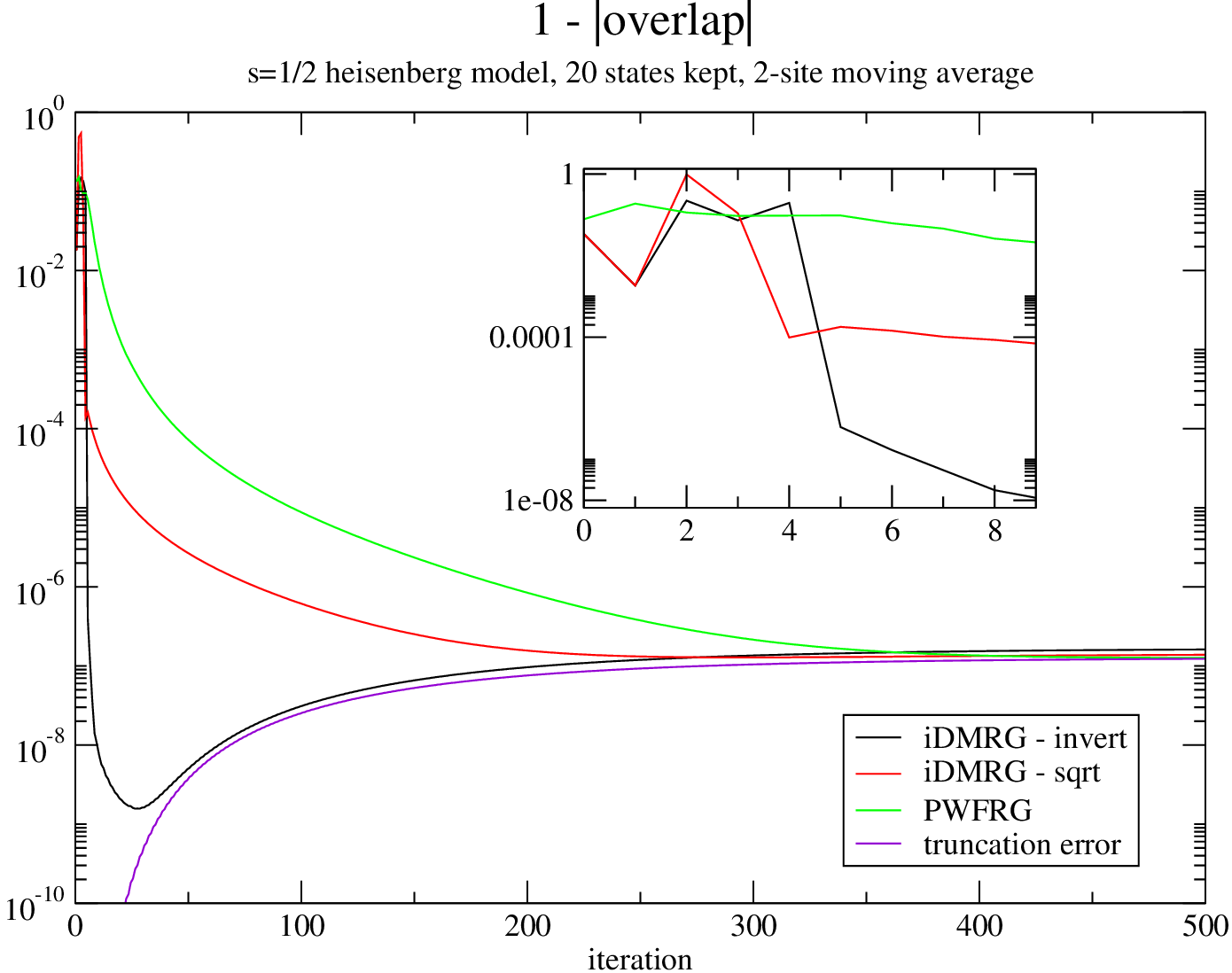} \par
\caption{Fidelity $1-\braket{\Psi_n^{\mathrm{trial}}}{\Psi_n}$ 
of the wavefunction transformation for iDMRG and PWFRG. For comparison, the truncation
error is also included, which sets a bound for the maximum possible fidelity.
To remove small period two oscillations
caused by the two-site unit cell, a two-site moving average was used for all quantities. }
\label{fig:Comparison}
\end{figure}

Figure \ref{fig:Comparison} shows the effectiveness of the iDMRG wavefunction transformation compared with
PWFRG for a simple example. This figure shows the fidelity of the initial wavefunction with respect 
to the variational minimum, $1-\braket{\Psi_n^{\mathrm{trial}}}{\Psi_n}$
for the spin-1/2 Heisenberg model with a 2-site unit cell and $m=20$ states kept in an $SU(2)$-symmetric basis 
\cite{NonAbelian,DMRGMPS}. 
The correlation length for this basis size is $\sim 48.5$ lattice sites. The exact groundstate of this
Hamiltonian is critical, however for a finite basis size the correlation length of a matrix product
wavefunction will always be finite. \cite{Fannes,Rommer} The signature of critical behavior is 
a scaling law for the correlation length as the basis size is increased, of the form $\xi = m^\kappa$,
where the exponent $\kappa \simeq 1.3$ for the Heisenberg model. \cite{Boman,Luca}
Also included in the figure is a variant
of \refeq{eq:iDMRGTransformation} that avoids the matrix inverse by taking instead the square root
of the singular values of $\Lambda^R_n \Lambda^L_n$ in the basis where $\Lambda_{n-1}$ is diagonal,
but this transformation is much less effective for
small systems. As expected, the PWFRG transformation converges only for system sizes much greater than
the correlation length due to the slow convergence of the translation operator.% (see the Appendix for details).

\begin{figure}[ht]
\centering
\includegraphics[width=\columnwidth,clip=true]{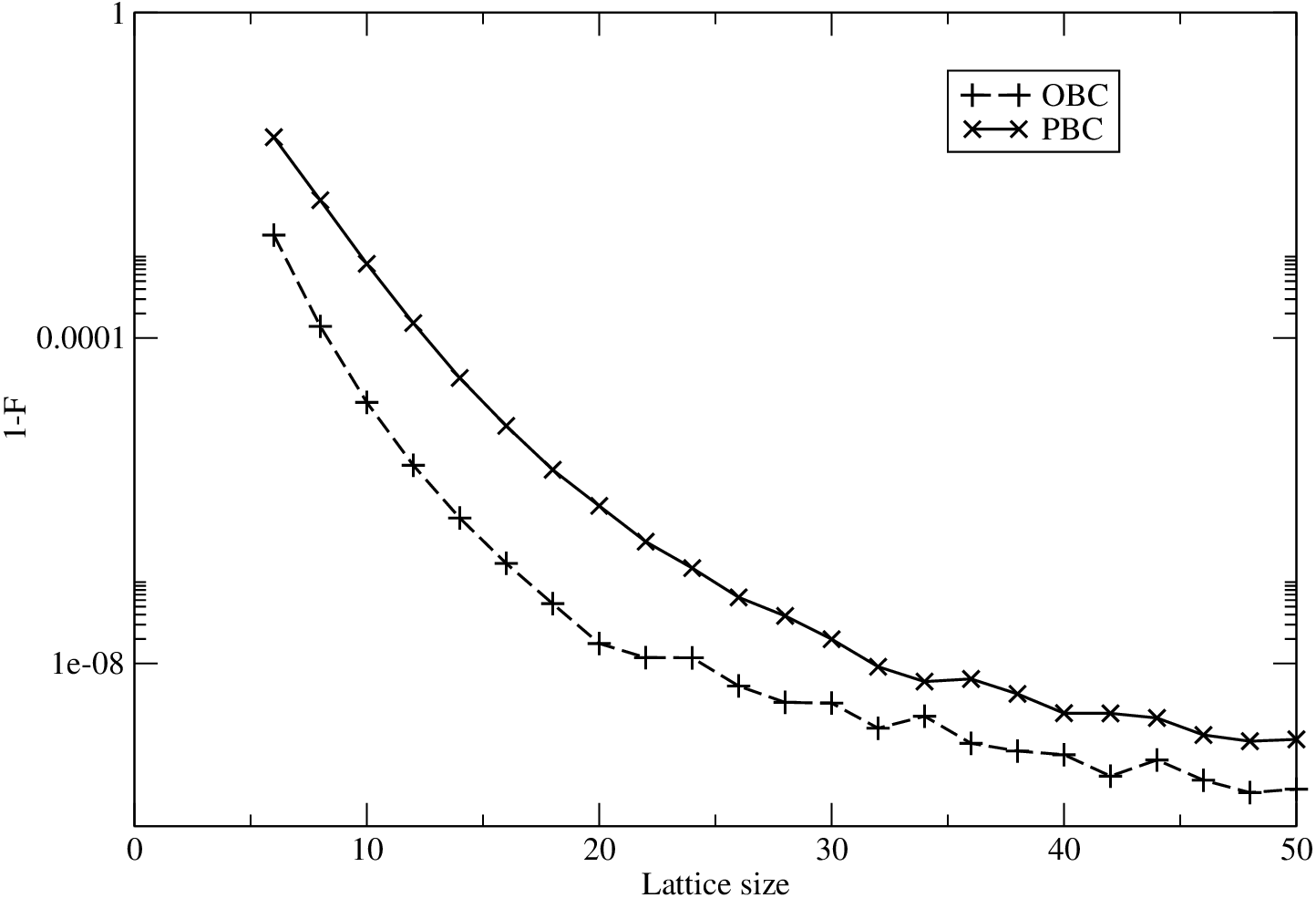} \par
\caption{Fidelity $1-\braket{\Psi_n^{\mathrm{trial}}}{\Psi_n}$ 
of the wavefunction transformation for open and periodic boundary conditions.}
\label{fig:OBCvsPBC}
\end{figure}

An interesting feature of \reffig{fig:Comparison} is the excellent performance of the iDMRG transformation
for systems much smaller than the correlation length where the transformation captures the effects of
the open boundary conditions extremely well, even for systems small enough to be exactly diagonalizable.
But for an exact diagonalization, periodic boundary conditions are more natural. So, in 
\reffig{fig:OBCvsPBC}, I compare the effectiveness of the transformation also for periodic 
boundary conditions, increasing the number of states kept to the limit of the numerical accuracy
of the wavefunction transformation. Using just 10 Lanczos iterations at each lattice size,
we obtain an energy for a 30-site system with periodic boundary conditions of
$E = -13.321963058(9)$, accurate to 10 significant figures, showing that this transformation could
be useful for accelerating exact diagonalization studies.

Clearly this algorithm has a similar function to iTEBD in the sense that both algorithms result in
a translationally-invariant variational approximation to the groundstate wavefunction.
But in fact the relationship between iDMRG and iTEBD is much deeper.
If we replace step 5 by an alternative,
\begin{equation}
\begin{minipage}{0.85\columnwidth}
\small
\begin{tabular}{cp{0.85\columnwidth}}
%\parbox{0.8\columnwidth}{
5. & (evolution) Obtain $\ket{\Psi_{n+1}}$ by applying a single bond evolution operator to 
sites $s'_{n+1}$ and $s_{n+1}$ of $\ket{\Psi^{\mathrm{trial}}_{n+1}}$,
\end{tabular}
\end{minipage}
\end{equation}
then the iTEBD algorithm is recovered \textit{exactly}, where the translationally invariant wavefunction
at the $n^{\mathrm{th}}$ step is given, in notation corresponding with that of Vidal\cite{iTEBD}, as,
\begin{equation}
\ldots \Lambda^{}_B \Gamma^{s_1}_A \Lambda^{}_A \Gamma^{s_2}_B 
\Lambda^{}_B \Gamma^{s_3}_A \Lambda^{}_A \Gamma^{s_4}_B \Lambda^{}_B \ldots \; ,
\end{equation}
where 
$\Gamma^{s}_A = \Lambda_{n-1}^{-1} A^s_n$, 
$\Lambda_A = \Lambda_n$,
$\Gamma^{s}_B = B_n^s \Lambda_{n-1}^{-1}$, and
$\Lambda_B = \Lambda_{n-1}$.
This is remarkable, because the
iTEBD algorithm was originally conceived as evolving, simultaneously, all bonds on an
infinite-size translationally invariant wavefunction, but it is formally equivalent
to a lattice-growth scheme where two sites are added to the lattice at each iteration
and the evolution is done performed via a single bond operator at the center of the lattice.
This latter interpretation has advantages, in particular the basis states in the iDMRG point
of view are always orthogonal, which provides a framework from which the convergence and stability
properties are amenable to analysis.

\subsection{The orthogonality fidelity}

A measure of how close the wavefunction is to the fixed point is obtained by comparing the wavefunctions
at successive iterations. At the fixed point, $\Lambda_n^R \Lambda_{n-1}^{-1}$ will be a diagonal
matrix that represents the phases of the eigenvectors chosen in the truncation procedure (if there
are degenerate eigenvalues then the corresponding block can be an arbitrary unitary matrix) that can
be incorporated into the neighboring matrices $A^{s'_{n+1}}_{n+1}$ so that the wavefunction 
transformation is the identity operation. Similarly, the diagonal matrix of phases of 
$\Lambda_{n-1}^{-1} \Lambda_n^L$ can be incorporated into $B^{s_{n+1}}_{n+1}$.
A measure for how similar the wavefunctions are can easily be obtained by calculating the fidelity 
between these wavefunctions at successive iterations, which I call the \textit{orthogonality fidelity},
\begin{equation}
F_{\mathrm{ortho}} = \braket{\Lambda_{n}}{\Lambda_{n-1}} \; ,
\end{equation}
 which is best obtained via the reduced density matrices,
\begin{equation}
F_{\mathrm{ortho}} = \mathrm{Tr} \sqrt{\sqrt{\rho^R_{n}} \; \rho_{n-1} \; \sqrt{\rho^R_{n}}} \; ,
\end{equation}
with the density matrices given by $\rho^R_{n} = \Lambda_{n}^{R\dagger} \Lambda_{n}^R$ and
$\rho_{n-1} = \Lambda_{n-1}^{\dagger} \Lambda_{n-1}^{}$.
This reduces to the sum of the singular values of $\Lambda_n^R \Lambda_{n-1}^\dagger$,
so is quite easy and efficient to calculate.
From the TEBD point of view, this measure shows how close
the basis states are to being orthogonal.
If there is any discarded weight in the truncation procedure of \reffig{fig:iDMRG} step 5,
this has a corresponding effect on $F_{\mathrm{ortho}}$. That is, in iDMRG the wavefunction
only reaches an exactly orthogonal fixed point if it is arranged that the truncation error
is zero, or in iTEBD the requirement for exact orthogonality is that the time-step goes to zero.

This points to an undesirable side-effect of a non-zero truncation of the wavefunction.
While having a finite truncation is generally highly beneficial to convergence compared with,
eg.~single-site algorithms with no density-matrix mixing that are highly susceptible to
convergence problems, near the fixed point of iDMRG a finite truncation error is a hindrance.
To investigate the effects of truncation errors on the fixed point, I have compared
iTEBD for various time-steps as well as
two-site and single-site iDMRG algorithms with respect to how close the resulting state is
to optimal, for a given number of states. Much of this section applies as well for finite-size
calculations, although probably with less extreme differences.

Compared with standard two-site (i)DMRG, the (i)TEBD algorithm
with a careful reduction in time-step towards zero,
produces a better wavefunction for a given number of states.
Surprisingly, this difference is quite substantial, in all cases I looked at amounting to around 30\% 
improvement in the number of states needed for a given accuracy. In most applications of DMRG this is 
of no consequence, and it is not an efficient use of computer resources to try to converge to
the best possible wavefunction for a given number of states. Instead of performing many many iterations
to get `optimal' convergence, often a better variational state is
achieved with a less-well converged wavefunction by a small increase in the basis size.
% (indeed, the message of this section
%is that achieving this ideal is quite expensive, even if it achievable at all).
Nevertheless in some cases the number of states is all-important. 
In particular, to calculate an optimal state
for a given $m$ in a scaling analysis, 
an error bar of 30\% is unacceptable. In addition, for applications such as time evolution,
where the number of states kept grows substantially (often exponentially) during the calculation, a 30\%
reduction in the number of states required for the initial state may lead to a substantial improvement
in the efficiency of the evolution.
In a time-evolution calculation, obtaining the initial state is usually much faster than the 
time evolution itself, 
so additional time spent on improving the initial state is worthwhile.

Takasaki, Hikihara and Nishino \cite{TakasakiFixedPointDMRG} have
proposed a combination of two-site and single-site algorithms, using the single-site algorithm in the
last few sweeps of the calculation. However, this still doesn't produce the best possible result because
the best states to keep do not correspond exactly with the largest $m$ eigenstates of the density matrix.
To be more precise, we can interpret the two-site algorithm as equivalent to a single-site algorithm with
an environment that is perturbed by additional degrees of freedom
corresponding to the second site. This perturbed environment means that the state 
isn't exactly translationally
invariant due to the appearance of a shallow bound-state at the central two sites, 
as noted by Dukelsky \textit{et al}. \cite{SierraVariational}
But this perturbation to the wavefunction is such that the 
largest $m$ density matrix
states do not give the optimal states to keep at the point of crossing over to 
the single-site algorithm. With no 
good quantum numbers this
is not such a serious problem and I would expect that the algorithm converges to 
a nearly-optimal state regardless (although possibly quite slowly). 
However practically all problems of interest utilize good quantum numbers, and in that case
the choice of the number of states in each symmetry sector,
which is fixed and does not change during the course of the single-site algorithm, 
will not be optimal. The solution is 
instead to evolve the wavefunction slowly,
such that as the fixed point is approached the change in wavefunction at each 
iteration smoothly goes to zero.
This does not happen in two-site DMRG because at the `fixed point' the change in wavefunction per iteration 
is precisely the truncation
error, which is usually small but non-negligible. A slow evolution of the wavefunction 
corresponds to 
imaginary time-evolution with a small time-step, which could be
done either with TEBD via the Trotter-Suzuki decomposition, or directly with DMRG by replacing the
eigensolver by a multiplication with $1-\delta\tau H$.
%Note however that
%there is no need to use a Trotter-Suzuki decomposition; the most straightforward modification 
%from standard DMRG
%is to evolve the wavefunction by $\ket{\psi}' = (1 - \Delta \tau H) \ket{\psi}$ which 
%avoids the complication
%of a finite Trotter error. 
A similar effect could be obtained simply by performing just 
a single iteration of
the eigensolver (Lanczos or similar) in the final sweeps of standard two-site DMRG.
An alternative approach that should achieve the same effect is White's single-site 
density matrix mixing approach
\cite{WhiteSingle}, and smoothly sending the density matrix perturbation to zero in the final
stages of the calculation.

\subsection{Explicit orthogonalization}

The previous section has argued that, regardless of which algorithm is used to obtain the
wavefunction, a finite truncation of the wavefunction leads a loss of orthogonality,
characterized by an orthogonality fidelity $1-F_{\mathrm{ortho}}$ which will be 
of the same magnitude as the discarded weight of the truncation.
% \footnote{Of course, a
%wavefunction that has not yet reached the fixed point will also be non-orthogonal.}
This is not ideal, since even a small non-orthogonality
affects the computation of expectation values, and the
iterative algorithms for expectation values described in \refsec{sec:Expectation} below
are simpler and perform better when the wavefunction is exactly orthogonalized. 
Fortunately, this presents
no problem as a procedure for explicitly orthogonalizing a translationally invariant MPS
has been developed by Or\'us and Vidal \cite{RomanTEBD}. 

For concreteness let us take a unit cell of 2 sites, so the wavefunction at some iteration $n$ 
is given by the matrices $A_n^{s'_n} \Lambda_n B_n^{s_n}$ together with the center matrix at the previous
iteration, $\Lambda_{n-1}$. Rotating the center matrix to the right gives the unit cell in
the form of left-orthogonalized matrices,
\begin{equation}
A_n^{s'_n} A^{s'_{n+1}}_{n+1} \Lambda^R_n \Lambda_{n-1}^{-1} \; ,
\end{equation}
and the infinite wavefunction comprises this set of matrices repeated. ie, with 
$P=\Lambda^R_n \Lambda_{n-1}^{-1}$ the wavefunction is
\begin{equation}
A_n^{s_1} A^{s_{2}}_{n+1} P A_n^{s_3} A^{s_4}_{n+1} P A_n^{s_5} A^{s_6}_{n+1} P \ldots \; .
\end{equation}
If the wavefunction is exactly orthogonal, then $P$ will be
the identity matrix (up to a unitary matrix that could be incorporated into one of the $A$-matrices),
and the orthogonality fidelity will be $1-F_{\mathrm{ortho}} = 0$.
But if $P \neq 1$, the MPS is not orthogonal because no matter whether we attach $P$ to the
$A$-matrix on the left or the right, the resulting $A$-matrix will not satisfy the orthogonality
constraint \refeq{eq:LeftOrtho}. A simple-minded way of making the MPS orthogonal would be to
simply rotate the $P$ matrices to the right iteratively, until a fixed point of the updated $A$-matrices
is reached. \footnote{This procedure is equivalent to running iterations of iDMRG/iTEBD with no update
to the wavefunction (ie. omitting step 5 of \reffig{fig:iDMRG}).} This is clearly not efficient, however.
Instead, the procedure of Or\'us and Vidal \cite{RomanTEBD} is to find a basis such that
the dominant eigenmatrix of the \textit{transfer operator} is the identity matrix \cite{Rommer}.
The transfer operator in the left-orthogonal basis $T_L$ is simply the identity operator of the unit cell, 
given by,
\begin{equation}
T_L(E) =  \sum_{s_1,s_2} P^\dagger A^{s_2\dagger}_{n+1} A_n^{s_1\dagger} E A_n^{s_1} A^{s_{2}}_{n+1} P \; .
\label{eq:LeftTransferOperator}
\end{equation}
For an orthogonal state, \refeq{eq:LeftOrtho} implies $T_L(1) = 1$ is a fixed point,
as we would expect for a representation of the identity operator.
For a non-orthogonal state we have instead $T_L(1) = P^\dagger P$.
So, in order to transform to a basis in which $P=1$, we obtain, with an efficient 
eigensolver such as the Arnoldi method \cite{GolubVanLoan}, the dominant eigenvector $V_L$ of $T_L$,
which we decompose into
\begin{equation}
V_L = X^\dagger X \; .
\label{eq:LeftDecomp}
\end{equation}
Note that $V_L$ is always Hermitian and non-negative. The factorization \refeq{eq:LeftDecomp}
could be done by a Cholesky facorization, however this is unstable if $V_L$ is close to singular 
\cite{GolubVanLoan};
factorizing via a singular value or eigenvalue decomposition is more robust.
We now perform a similarity transformation on the unit cell,
\begin{equation}
A_n^{s_1} A^{s_{2}}_{n+1} P \rightarrow X A_n^{s_1} A^{s_{2}}_{n+1} P X^{-1} \; ,
\end{equation}
and in this new basis, $T'_L(1) = 1$ as required. Note that to obtain the new $\Lambda$ matrices,
an additional step is needed. This could be done by solving the fixed point relation
\refeq{eq:FixedPointDM} but here we follow the outline of \refonline{RomanTEBD} and solve the corresponding
right-orthogonal transfer matrix for the right-orthogonalized unit cell,
\begin{equation}
T_R(E) =  \sum_{s_1,s_2} Q B^{s_1}_{n} B_{n+1}^{s_2} E 
B_{n+1}^{s_2\dagger} B^{s_1\dagger}_{n} Q^\dagger \; ,
\label{eq:RightTransferOperator}
\end{equation}
where $Q = \Lambda_{n-1}^{-1} \Lambda^L_n$. Similarly to the left eigenvector, we now obtain the
eigenvector $V_R$ of $T_R$, and decompose into
\begin{equation}
V_L = Y Y^\dagger \; ,
\end{equation}
from which we see that the complete set of transformations is,
\begin{equation}
\begin{array}{rcl}
A_n^{s_1} &\rightarrow& X A_n^{s_1} \\
B_{n+1}^{s_2} &\rightarrow& Y B_{n+1}^{s_2} \\
\Lambda_{n-1} &\rightarrow& X \Lambda_{n-1} Y \\
\Lambda^R_n &\rightarrow& \Lambda^R_n Y \\
\Lambda^L_n &\rightarrow& X \Lambda^L_n \\
\end{array}
\end{equation}
This choice of tensors has the advantage over that used in \refonline{RomanTEBD} that the 
inverses of the $X,Y$ matrices are not actually required, only the inverse of $X \Lambda_{n-1} Y$
which determines the singular values in the orthogonal basis. There is one other difference
here to the procedure described by Orus and Vidal, concerning the treatment
of unit cells of $N>1$ sites. Orus and Vidal suggested handling these unit cells with
a coarse-graining procedure that combines
a unit cell of $N$ cites into a single site with an increased local dimension. This is inefficient,
as it corresponds to evaluating the summations in the eigenvalue equation \refeq{eq:LeftTransferOperator}
in a non-optimal way, using $O(d^N)$ matrix multiplies rather than $O(Nd)$ multiplies if the summations
are performed sequentially. Aside from this difference, and the slightly different choice of tensors,
the procedure described here is equivalent to that of Or\'us and Vidal.
Note that this procedure orthogonalizes the unit cell, it does not however guarantee that
the individual sites comprising a multi-site unit cell separately satisfy \refeq{eq:LeftOrtho}.
The step of orthogonalizing each site should be done for the sake of numerical robustness,
and is readily achieved through the usual orthogonalization procedure,
described in \refsec{sec:MPS}.

The orthogonalization procedure is quite efficient, and for robustness I always perform this
orthogonalization step before calculating the final expectation values.

\section{Expectation values}
\label{sec:Expectation}

In this section, we consider expectation values of MPO's, firstly for the case of an operator that acts
only on a finite range of the lattice, then for translationally invariant
triangular MPO's. 

\subsection{Finite-range operators}

Any finite-range operator of width $N$ can be expressed as a position-dependent MPO with
matrices $W_{1} \ldots W_{N}$ with an open-boundary condition 
representation. That is, the left-hand MPO matrix is a $1\times M$ row vector, and the right-hand MPO
is a $M' \times 1$ column vector. Starting from the right-hand side, the initial operator represents
the identity operator on an infinite lattice, which will be trivial if the state is orthonormalized.
Thus define $F^1_0 = 1$ to be this identity operator, then starting from the right hand side,
\begin{equation}
F_{n+1}^{\alpha'} = \sum_{\alpha,s',s} W_{(n+1)\alpha'\alpha}^{s's} B^{s'} F_n^\alpha B^{s\dagger} \; .
\end{equation}
The final iteration results in a single matrix $F_N^1$ that represents the operator matrix elements.
The expectation value is then given by $\Tr F_N^1 \rho$ as usual, where $\rho$ is the 
density matrix for the
semi-infinite strip, being the fixed point of $\rho = \sum_s B^{s\dagger} \rho B^s$.

\subsection{Triangular operators}
\label{sec:ExpectationTriangular}

To evaluate an expectation value of a lower-triangular MPO, we need to determine the matrix elements as a 
function
of the wavefunction $A$-matrix elements. For an MPO of dimension $M \times M$, we have $M$ different matrices
$E^\alpha$, $\alpha = 1,2,\ldots,M$, that represent the components of the MPO on the remainder of the lattice.
For concreteness, we firstly assume that the operator is of the form of a sum of short-range terms.
This means that the diagonal of the MPO is zero, except for the top-left and bottom-right entries which
are equal to the identity operator. This gives the fixed-point recursion
\begin{equation}
E^{\alpha'} = \sum_{\alpha,s',s} W^{s's}_{\alpha'\alpha} A^{s'\dagger} E^\alpha A^s - E_0 \delta_{\alpha',1} \; ,
\label{eq:EMatrixFixedPoint}
\end{equation}
where $E_0$ is the expectation value per site.
This is straightforward to evaluate term by term. Starting from the last entry at $\alpha=M$, since
the bottom-right entry is the local identity operator we have,
\begin{equation}
E^M = \sum_s A^{s\dagger} E^M A^s \; ,
\end{equation}
which is just the fixed point of the orthogonalization constraint, so for left-normalized matrices
$E^M = 1$ identically.
The remaining matrices $E^\alpha$ for $1 < \alpha < M$ are simple to construct in descending order
because these matrices
are functions only of the previous matrices $E^{\alpha'}$ for $\alpha' > \alpha$.
The exception is the final matrix $E^1$, which again has an identity operator on the diagonal, giving
the linear equations
\begin{equation}
E^1 - \sum_{s} A^{s'\dagger} E^1 A^s + E_0 = \sum_{\alpha>1,s',s} 
W^{s's}_{1\alpha} A^{s'\dagger} E^\alpha A^s \; ,
\end{equation}
where the expectation value $E_0$ is given by the trace of the right hand side.
If only the expectation value is required, these linear equations do not need to be actually solved.
The full solution is required only if the matrix $E^1$ needed, for example to restart an iDMRG calculation
from a given translationally invariant wavefunction.

The procedure when there are additional matrix elements on the diagonal of the MPO follows similarly. 

% In general, we do not require the operator to be translationally invariant, and it is straightforward to 
% generalize
% to operators that are non-zero eigenstates of translation. This implies a (unitary) matrix 
% representation of the
% translation operator, such that a translation of an MPO by one site, $T(M)$ (acting only on the 
% local basis of $M$), 
% is given by
% \begin{equation}
% T(M) = U_T M U_T^{\dagger} \; ,
% \end{equation}
% where $U_T$ acts only in the matrix basis of $M$. 
% For example, a particle creation operator with finite momentum $k$ is represented as
% \begin{equation}
% M = \begin{pmatrix} 
%   I & 0 \\ 
%   c^\dagger & e^{ik} I
% \end{pmatrix} \; ,
% \end{equation}
% with the translation operator
% \begin{equation}
% U_T = \begin{pmatrix} 
%   I & 0 \\ 
%   0 & e^{ik} I
% \end{pmatrix} \; .
% \end{equation}

\subsection{Transfer matrix and symmetry breaking}

The spectrum of the transfer matrix \refeq{eq:LeftTransferOperator} is very useful as this
determines the possible correlation lengths in the system. \cite{RommerOstlund}
The transfer matrix has always at least one eigenvalue equal to 1, as follows from
the normalization constraint \refeq{eq:LeftOrtho}. The next leading eigenvalue determines
the longest correlation length in the system. Normally this eigenvalue will be strictly
less than 1. Even in the presence of long-range order, a non-zero order parameter
will lead to a spontaneously broken symmetry at the infinite-size fixed point, as long
as the symmetries that are imposed on the MPS representation allow this. 
In an infinite size system, a broken symmetry implies multiple groundstates which 
are exactly degenerate. Such spontaneous symmetry breaking typically does not occur in
finite-sized systems because there are usually matrix elements connecting the
groundstates that lift the degeneracy, and such matrix elements are exponentially
small in the lattice size $\sim \exp[-L]$. Note that translation symmetry is an exception to this,
since a finite system inherently breaks translation symmetry from the beginning.

Spontaneous symmetry breaking of \textit{discrete} symmetries 
is realized in an infinite MPS with matrix elements connecting degenerate
groundstates going to exactly zero at the fixed point. Hence we expect in a numerical
simulation that exactly one of the degenerate states will be chosen non-deterministically.
(If any of the groundstates has a lower entanglement then this state would tend to be preferentially selected
by the variational principle, as this state would have a lower energy for a particular basis size.)
%This is the case for example in a ferromagnet where the maximal $S^z$ state has lower entropy
%than the $S^z=0$ state of the same total spin.)
However, if good quantum numbers are used then the obtained groundstate will always have this symmetry
even if it would otherwise be spontaneously broken. For example, 
In the low-field ordered phase of the ITF model \refeq{eq:ITF}, there are two degenerate thermodynamic
groundstates, namely the two fully polarized states $\ket{\downarrow \downarrow \ldots}$ and
$\ket{\uparrow \uparrow \ldots}$. In the absence of any mechanism enforcing $Z_2$ symmetry,
iTEBD or iDMRG will spontaneously select one of these groundstates, selected eg.~by numerical
noise from finite-precision arithmetic that leads to one groundstate having slightly higher
weight than the other, which then quickly dominates. \footnote{This is assuming that the number
of states kept is not so big that multiple degenerate groundstates can be represented with
high fidelity. In that situation the stability of the algorithm is in question,
eg.~it is possible that a small numerical noise could result in multiple copies 
of the \textit{same} groundstate, thus leading to a catastrophically non-orthogonal basis.}
However if a $Z_2$ quantum number is used, then
the obtained groundstate will be an exact eigenstate of $Z_2$, namely
$\ket{\downarrow \downarrow \ldots} \pm \ket{\uparrow \uparrow \ldots}$, which is a
GHZ state \cite{GHZ} with long-range order leading to an additional eigenvalue of 
unity in the transfer matrix spectrum.
Thus, the transfer matrix is very useful for diagnosing the presence of long-range order, without
\textit{a priori} knowledge of the specific order parameter.
Similar considerations also apply for the transfer matrix extended to string order
parameters, \cite{VerstraeteString} where the spontaneous symmetry breaking is instead associated
with edge excitations that do not affect the bulk wavefunction \cite{NijsKennedy}.

\begin{figure}
\centering
\includegraphics[width=\columnwidth,clip=true]{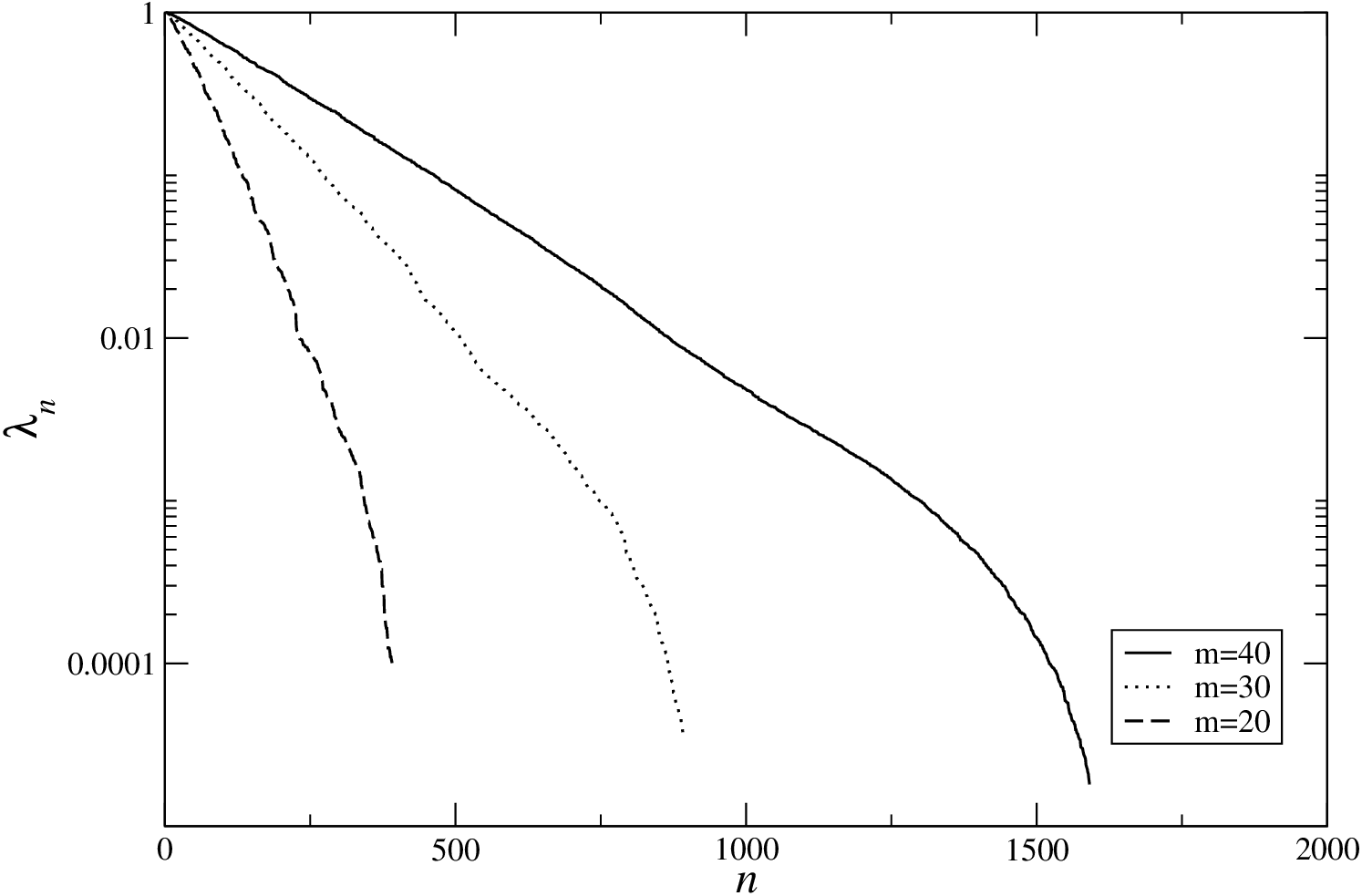} \par
\caption{The spectrum of the transfer matrix for the ITF model near the critical point.
The next-leading eigenvalue determines the correlation length, here
$\xi=578.7,1236.6,2539.2$ for $m=20,30,40$ respectively.}
\label{fig:IsingTransferSpectrum}
\end{figure}

Figure \ref{fig:IsingTransferSpectrum} shows the transfer matrix spectrum of the ITF model in the vicinity
of the critical point $\lambda=1$, for some different basis sizes. As can be seen, the ITF
model, which is remarkable for having a very large correlation length even with
a very small basis size, has relatively few eigenvalues that contribute to long range
correlations. For example, considering only the eigenvalues corresponding to correlation lengths
of greater than 10 sites gives only around 30 relevant eigenvalues, even for the case of $m=40$ where the
correlation length is a few thousand sites. Thus the long-distance behavior of correlation functions 
can be determined by looking at just a few eigenvalues.
%For a more typical example, a slightly
%frustrated spin 1/2 $J_1-J_2$ (zig-zag) spin chain with $J_2 / J_1 = 0.2$,
%with $m=1000$ I obtain a correlation length of $\sim 400$ and around 60 eigenvalues have a correlation
%length in excess of 10 sites. ??? That cannot be true, that xi=400 with as m=1000.
In the case of a critical system, 
the power-law behavior of a correlation function is obtained approximately, as the sum of
a set of exponentials \cite{RommerOstlund}. Hence the eigenvalues of the transfer matrix
give a good guide as to the range of validity where an approximate power-law can be expected,
ie. to a distance $\sim$ the maximum correlation length in the system.
A complication of using the transfer matrix directly to expand a correlation function as
a set of exponentials is that the transfer matrix of \refeq{eq:LeftTransferOperator} is
not a normal matrix and hence the eigenvectors are not orthogonal. But this
presents no essential difficulty, as long as care is taken to project out the
large number of components with short correlation lengths prior to determining
the amplitudes (expansion coefficients) of each mode, eg.~by calculating the
operator $E$-matrix for a correlation at some distance larger than the smallest 
correlation length eigenvalue obtained from the transfer operator.

\subsection{Fidelity}

The overlap of two MPS is a useful operation that is used to obtain the \textit{groundstate fidelity}
$F(g_1,g_2) = |\braket{g_1}{g_2}|$ between groundstates of different interaction parameters $g_1$ and $g_2$.
The fidelity has attracted much recent attention as a tool for probing quantum phase transitions. \cite{Fidelity}
For an MPS, this quantity is always of the form $d^L$, for an $L$-site lattice with $0 < d < 1$. Thus, the fidelity
is measured per site in much the same way as the expectation value of an exponential MPO (indeed, the fidelity
is equivalent to the expectation value of the identity operator which can be seen as the simplest possible exponential MPO).
The \textit{fidelity susceptibility} \cite{YouFidelity} is a measure of the curvature of the fidelity
at a particular point, and hence shows how quickly the wavefunction is changing with the interaction parameter.
The fidelity susceptibility per site is given by,
\begin{equation}
\chi_{F}(g) = \lim_{\delta g \rightarrow 0, L \rightarrow \infty} \frac{-2 \ln F(g,g+\delta g)}{L \; \delta g^2} \; .
\end{equation}
For finite MPS, the fidelity has been proposed as a measure since at least the work of Andersson, Boman and \"Ostlund. \cite{Boman}
For an infinite MPS, the fidelity (and hence also the fidelity susceptibility) is straightforward to calculate
via an eigenvalue equation. Similar to the transfer operator (which could be seen as calculating the norm $\braket{\Psi}{\Psi}$),
the fidelity between two states $\ket{\Psi{1}}$ and $\ket{\Psi{2}}$ is given by the largest eigenvalue of an
overlap operator, here given for a two-site unit cell example (compare
\refeq{eq:LeftTransferOperator}, assuming the states are orthogonalized),
\begin{equation}
T_{12}(E) =  \sum_{s_1,s_2} A^{s_2\dagger}_{2}(1) A_1^{s_1\dagger}(1) E A_1^{s_1}(2) A^{s_{2}}_{2}(2) \; ,
\end{equation}
where $A_n^{s_n}(1)$ and $A_n^{s_n}(2)$ are the matrices representing $\ket{\Psi{1}}$ and $\ket{\Psi{2}}$ respectively.
The fidelity per unit cell is then given by the eigenvalue $\lambda$ of $T_{12}(E) = \lambda E$, which can be any complex number
of magnitude $\leq 1$.

In a finite MPS, the fidelity is sometimes complicated by boundary effects that are irrelevant to the bulk properties
of the wavefunction. For a concrete example, the bilinear-biquadratic $S=1$ chain \cite{Spin1Chain} has a variety of groundstate phases
including a gapped resonating valence bond (RVB) phase (including the AKLT point \cite{AKLT} that is exactly solved with
a non-Abelian MPS with dimension $m=1$), a gapped dimer phase, and a critical trimerized phase. The RVB phase is characterized
by effective spin-$1/2$ edge excitations that are best treated by applying \textit{real} spin-$1/2$ states at the boundaries,
which means that the bulk wavefunction in the infinite size limit has quantum numbers that are all half-integer. This presents
no problems for MPS algorithms in the infinite limit, however in other phases such as the gapped dimer phase it is best
\textit{not} to supply spin-$1/2$ states at the boundaries, as these frustrate the formation of dimer pairs and
also introduces spurious effects into the fidelity. On the other hand, to calculate the fidelity for a finite size system the 
Hilbert space of the wavefunctions must coincide exactly, which causes severe difficulties in calculating the fidelity
across the RVB-dimer phase transition.
These problems do not arise in the infinite MPS case, as the boundary effects are not present and having different spin representations
occurring in the bond basis of the wavefunctions is no problem for the fidelity calculation.

\section{Conclusions}
\label{sec:Conclusions}

In this paper I have examined the infinite-size DMRG algorithm from the perspective of the matrix product
approach, which gives directly a translationally invariant fixed point previously studied by
Rommer and \"Ostlund \cite{Rommer,RommerOstlund}. The core development in this algorithm is the
wavefunction transformation, which improves upon that used in PWFRG \cite{PWFRG}, and is useful
even for small finite systems. \refsec{sec:Expectation} has shown how to calculate expectation values
in the thermodynamic limit of operators directly from the matrix product form of an observable,
which gives access to higher moments such as the energy variance. The spectrum of the transfer matrix
gives detailed information on the correlation length and ordered phases, and expanding a correlation
function as a sum of exponentials corresponding to the modes of the transfer matrix
gives an alternative approach to fitting a critical exponent.
The groundstate fidelity between different groundstates is readily calculated, and this gives
a precise way to determine the fidelity that is free of complications due to boundary
effects.

Infinite-size DMRG represents an efficient method for finding translationally invariant states that
could replace many of the current uses for finite-size DMRG. However one situation in which
the current algorithms are not a sufficient replacement is the calculation of
excited states, for example energy gaps, low-lying excited states and spectral functions.
Despite some suggestions to the contrary \cite{WhitePBC},
I believe these calculations should be readily achievable. Indeed, the basic approach of representing
a particle excitation as a superposition of local perturbations with an MPS has already been discussed 
by Rommer and \"Ostlund\cite{RommerOstlund}, leading to an ansatz for an excitation,
for example the application of a particle creation $c^\dagger_k$ at some
momentum $k$ to an infinite MPS. Building upon this approach 
I see no reason why there should not exist efficient
infinite-size counterparts for the well-established correction vector \cite{CV} and dynamical DMRG \cite{DDMRG} 
algorithms for frequency-space dynamical correlations of finite-size systems.
Note that calculating spectral functions via real-time evolution of a local particle excitation 
\cite{WhiteTime}
is already achievable, by using a section of a
translationally invariant state as the input for a finite-size time evolution.

The iDMRG algorithm presents an interesting dichotomy of solving simultaneously for both the 
wavefunction $A$-matrix elements and the Hamiltonian $E$-matrix elements.
The present scheme uses an efficient Lanczos eigensolver to obtain the wavefunction,
while the effect of growing the lattice one site at a time amounts to
using the less efficient power method to obtain the Hamiltonian matrix elements,
in the sense that exactly one iteration of the $E$-matrix multiply is performed per step.
Therefore, this algorithm will converge in a minimum number of iterations
that is at least as large as the correlation length. On the other hand,
\refsec{sec:ExpectationTriangular} shows how to obtain the Hamiltonian matrix elements of a
translationally invariant state much more efficiently using a linear solver. This suggests
an accelerated algorithm that efficiently solves simultaneously the fixed point wavefunction and the
associated fixed point Hamiltonian \refeq{eq:EMatrixFixedPoint}, which could in principle 
converge in a number of iterations much smaller than
the correlation length at the cost of the algorithm being non-linear. 

Algorithms for finding the groundstate of a one-dimensional Hamiltonian form a component
of some algorithms for higher dimensional systems, most notably the iPEPS algorithm 
\cite{iPEPS} for 2D tensor product states \cite{Maeshima}. The higher efficiency of
direct eigensolvers should serve a similar function as the 1D case, although the analogue
of a triangular MPO for the Hamiltonian construction is clearly more complicated in 2D.
\cite{Crosswhite}

\begin{acknowledgments}
Many thanks to Ulrich Schollw\"ock, Thomas Barthel, Guifre Vidal and Luca Tagliacozzo
for stimulating conversations, and even more thanks to Ulrich and Luca for 
providing valuable comments on the manuscript.
\end{acknowledgments}

% \appendix*
% \section{PWFRG}
% \label{sec:PWFRG}

% This appendix describes briefly the PWFRG algorithm \cite{PWFRG,PWFRG2,PWFRG3,PWFRG-MPS} and the relationship to
% the iTEBD and iDMRG algorithms.

% \begin{equation}
% R^s = \sum_s A^{s\dagger} R^s A^s
% \end{equation}
% For simplicity, we take a unit cell of one site, 
% and we treat only the left-orthogonalized matrices at the $n^{\mathrm{th}$ iteration,
% $A_n^{s_n}$, eg, take the right-orthogonalized states to be the parity reflection.
% The PWFRG transformation 

\end{document}